# Universal clustering structure and $C \approx 0.85$ scaling in complex earthquake networks

Sumiyoshi Abe[a], Norikazu Suzuki[b]

[a] *Department of Physical Engineering, Mie University, Mie 514-8507, Japan*

[b] *College of Science and Technology, Nihon University, Chiba 274-8501, Japan*

**Abstract.** Earthquake network captures the complexity of seismicity in a peculiar manner. Given a seismic data, the procedure of constructing an earthquake network proposed in [S. Abe, N, Suzuki, Europhys. Lett. 65 (2004) 581] contains as a single parameter the size of the cells, into which a geographical region under consideration is divided. Then, the characteristics of the network depend on the cell size, in general. Here, the dependency of the clustering coefficient, *C*, of network on the cell size is studied. Remarkably, *C* of the earthquake networks constructed from the seismic data taken from California, Japan and Iran well coincide for each value of the rescaled dimensionless cell size. It is found that the networks in California and Japan are three-dimensional, whereas the one in Iran is rather two-dimensional. In addition, the values of *C* of all these three networks monotonically converge to $C \approx 0.85$ as the dimensionless cell size becomes larger than a common value, highlighting a universal aspect of the concept of earthquake network. An implication of the result to understanding the physics of seismicity is discussed.

1. **Introduction**

The concept of earthquake network has been introduced in Ref. [1] in order to describe the complexity of seismicity in an efficient way. Since then, a number of its remarkable properties have been revealed. It has been found for example that an earthquake network is scale-free [1,2], small-world [3] and hierarchically organized [4], and has a nontrivial community structure [5]. This network approach has opened a new possibility of extracting information about yet largely unknown dynamics governing seismicity.

The method of constructing an earthquake network proposed in Ref. [1] is as follows. Firstly, a geographical region under consideration is divided into cubic cells with the size *L*, i.e., the length of the side of the cell. A cell is regarded as a vertex if earthquakes with any values of magnitude occurred therein. Then, two vertices associated with two successive events are connected by an edge, which represents the event-event correlation (see the discussion in the paragraph after next). In particular, if two successive events occurred in the same cell, then a tadpole, i.e., a self-loop, is attached to that vertex. In this way, a given seismic data can be mapped to a growing stochastic network.

A convenient method for practically setting up the cells and identifying a cell for each earthquake is as follows. Let $\theta_0$ and $\theta_{max}$ be the minimal and maximal values of latitude of the whole region and $\phi_0$ and $\phi_{max}$ be the minimal and maximal values of longitude, respectively. These angles are measured in the unit of radian. We define $\theta_{av}$ as the sum of the values of latitude of all the events divided by the number of events. The hypocenter of the $i$-th event is denoted by $(\theta_i, \phi_i, z_i)$, where $\theta_i$, $\phi_i$ and $z_i$ are the values of latitude, longitude and depth, respectively. The north-south distance and the east-west distance between $(\theta_0, \phi_0)$ and $(\theta_i, \phi_i)$ are respectively given by $d_i^{NS} = R \cdot (\theta_i - \theta_0)$ and $d_i^{EW} = R \cdot (\phi_i - \phi_0) \cdot \cos\theta_{av}$, where $R\ (\cong 6370\text{ km})$ is the radius of the Earth. The depth is just $d_i^{D} = z_i$, as it is. Starting from the point $(\theta_0, \phi_0, z_0 \equiv 0)$, divide the region into cubic cells with a given value of the cell size. Then, the cell of the $i$-th event is determined in terms of $d_i^{NS}$, $d_i^{EW}$ and $d_i^{D}$.

The above-mentioned procedure of constructing a network is based on the working hypothesis that successive events are correlated *at least at the statistical level*. This hypothesis comes from the following empirical facts: (i) an earthquake can trigger the next far-off one [6,7], and (ii) the distributions of spatial distance and time interval between two successive events significantly deviate from the Poissonian [8-11]. These may allow us to frame the working hypothesis. Also, an event may trigger not only the next one but also the one after next etc.. Such a "non-Markovian" formulation is actually possible, but we avoid extra complication like that, here.

Previously, earthquake networks have been studied only for limited geographical regions, but, recently, an attempt has been made in Ref. [12] to construct and analyze the network of worldwide seismicity with magnitude $M \geq 4.5$. It is however noted that if the above-mentioned method of setting up cells is directly used for the whole globe, an obvious difficulty appears in the polar regions.

Clearly, a full earthquake network is a directed one with multiple edges as well as tadpoles. When its small-worldness is studied, it should be reduced to a simple network, where the directedness is ignored, tadpoles are removed, and each multiple edge is replaced with a single edge.

The method of earthquake-network construction explained above contains a single parameter, which is the cell size, $L$. Quantities characterizing a network depend on this parameter, in general. Since there is no *a priori* principle for determining the cell size, it is of crucial importance to clarify how change of the cell size affects the network characteristics. This problem has been studied in a recent work [13] (see also [14]). Among various network characteristics, we have studied in detail a *short-time behavior* of the clustering coefficient (see the next section), which is known to play an important role in specifying both main shocks and aftershocks [15]. It has been found [13] that, as a *long-time behavior*, the clustering coefficient, $C$ (its definition to be given in the next section), converges to a large specific value

$$C \approx 0.85, \qquad (1)$$

as the dimensionless cell size increases. Since all characteristic quantities of a network do not possess physical dimensions, the cell size should be made dimensionless. In the work [13], the following two different definitions of the dimensionless cell size have been considered:

$$l_2 = L / (L_{\text{LAT}} L_{\text{LON}})^{1/2}, \tag{2}$$

$$l_3 = L / (L_{\text{LAT}} L_{\text{LON}} L_{\text{DEP}})^{1/3}, \tag{3}$$

where $L_{\text{LAT}}$, $L_{\text{LON}}$, and $L_{\text{DEP}}$ are the dimensions of the whole geographical region under consideration in the directions of latitude, longitude, and depth, respectively. One could consider a more general form such as $L / (L_{\text{LAT}}^{\alpha} L_{\text{LON}}^{\beta} L_{\text{DEP}}^{\gamma})^{1/(\alpha+\beta+\gamma)}$, where $\alpha$, $\beta$, and $\gamma$ are nonnegative parameters with at least one of them being nonzero, but we do not introduce such additional parameters, here.

The seismic data analyzed in Ref. [13] are of California, Japan and Iran. The values of $C$ of the earthquake networks in these regions converge to the universal one in Eq. (1) as $l_2$ and $l_3$ increase. However, for small values of $l_2$ and $l_3$, $C$ of the network in Iran significantly deviates from those in California and Japan.

In this short note, we reexamine this problem concerning the cell-size dependence of the clustering coefficient of earthquake network. We will show that the networks in California and Japan are three-dimensional, whereas the one in Iran is quite two-dimensional, and an impressive universal behavior emerges if this issue of dimensionality is appropriately taken into account.

Throughout the analysis, no threshold is set on magnitude and all events recorded in the data sets are taken. This is due to the fact [4] that the hierarchical structure of network is supported by weak earthquakes. Then, one might care about limited detection ability, in reality. Regarding this point, however, one should recall [16] that complex networks are highly tolerant to random failures.

## 2. Universal clustering structure

To calculate the value of the clustering coefficient, we reduce a full earthquake to a simple network in the small-world picture, as mentioned in the preceding section. Such a network is characterized by a symmetric adjacency matrix, $A$, having the property: $(A)_{ij} = 1\,(0)$ if the $i$-th and $j$-th vertices are connected (unconnected). The clustering coefficient, $C$, is then expressed as follows:

$$C = \frac{1}{N} \sum_{i=1}^{N} c_i \qquad (4)$$

with

$$c_i = \frac{(A^3)_{ii}/2}{k_i(k_i-1)/2}, \qquad (5)$$

where $N$ and $k_i$ are the total number of vertices and the value of connectivity, i.e., the number of edges, of the $i$-th vertex, respectively. $C$ ranges from 0 to 1 and measures the tendency of triangle formation in the network, as can be seen in the $A^3$-structure in Eq. (5). The larger $C$ is, the more clustered the network is. One of the important features of a small-world network is that such a network has a large value of $C$ compared to the Erdös-Rényi classical random network [17].

We have constructed the earthquake networks by taking three different data sets of (i)

California; available at http://www.data.scec.org, (ii) Japan; at http://www.hinet.bosai.go.jp and (iii) Iran; at http://irsc.ut.ac.ir/. The periods and the geographical regions covered are as follows: (i) between 00:06:28.31 on 1 January 2005 and 13:52:11.55 on 11 November 2006, 28.00°N–39.41°N latitude, 112.10°W–123.54°W longitude with the maximal depth 35.00 km, (ii) between 00:02:52.93 on 1 June 2007 and 05:43:19.30 on 27 July 2007, 20.26°N–49.31°N latitude, 120.55°E–155.11°E longitude with the maximal depth 552.00 km, and (iii) between 03:08:11.10 on January 1, 2006 and 18:26:21.90 on December 31, 2008, 23.89°N–43.51°N latitude, 41.32°E–68.93°E longitude with the maximal depth 36.00 km, respectively. The total numbers of events in these periods are adjusted to be 22 845. This adjustment enables us to avoid the effect of data-size dependence [18]. As mentioned in Sec. 1, we do not set threshold on the value of magnitude and take all events included in the data sets.

We wish to stress the following. Seismicity in Iran occurs in a relatively shallow region compared to those in Californian and Japan. In fact, we have ascertained that 80% of the events occurred in the region shallower than $d=13.70$ km in California, $d=40.70$ km in Japan and $d=20.50$ km in Iran. On the other hand, the values of the corresponding "epicenter region", $\lambda$ (the square root of the product of the dimensions in the directions of latitude and longitude covering 80% of the events), are 1153.51 km in California, 3171.84 km in Japan and 2348.86 km in Iran. Therefore, the values of $d/\lambda$ are 0.012 in California, 0.013 in Japan and 0.009 in Iran, showing that seismicity in Iran is, in fact, shallower compared to those in California and Japan.

Taking this fact into account, we come back to the values of $(L_{LAT}L_{LON})^{1/2}$ and $(L_{LAT}L_{LON}L_{DEP})^{1/3}$ in Eqs. (2) and (3). They are respectively as follows: (i) 1153.51 km and 359.78 km, (ii) 3171.84 km and 1770.87 km and (iii) 2245.73 km and 566.25 km.

Now, we examine the dependency of the clustering coefficient, $C$, on the dimensionless cell size.

The result obtained is as follows. For comparison, we present Fig. 1 (see also Ref. [13]). There, we plot $C$ with respect to $l_3$ in Eq. (3). All three curves of $C$ approach the common value in Eq. (1). It is seen, however, that the curves of California and Japan coincide quite well, whereas that of Iran deviates from them for smaller values of $l_3$. On the other hand, now taking into account the fact that seismicity in Iran is rather two-dimensional, we present in Fig. 2 the plots of $C$ with respect to $l = l_3$ for California and Japan but $l = l_2$ for Iran. There, one observes a remarkable behavior that all three curves nicely collapse to a single one. Therefore, for the present data size, the scale of "coarse graining" is fixed as

$$l \approx 0.1, \qquad (6)$$

with the definition of $l$ itself being dependent on the dimensionality characterizing seismicity in each geographical region under consideration. This is our main result and highlights a universal aspect of the concept of earthquake network.

Closing this section, we would like to mention the following point. Actually, we have also examined the finite data-size scaling [18] by changing the periods and numbers of events in the three data sets. Consequently, we have ascertained that data collapse as in Fig. 2 cannot be realized as long as $l_2$ and $l_3$ are commonly used for the three data sets. Thus, the use of $l = l_3$ for the data from California and Japan and $l = l_2$ for that from Iran has turned out to be essential.

3.  **Conclusion**

We have reexamined the dependency of the clustering coefficient, $C$, of earthquake network on the dimensionless cell size, which is the one and only parameter contained in the network construction. Analyzing the networks constructed from the data sets taken from California, Japan, and Iran, we have found that $C$ exhibits a universal behavior and approaches $C \approx 0.85$ as the dimensionless cell size becomes larger than a certain common value (about 0.1 in the present case). This discovery has been made based on the fact that seismicity in Iran is rather two-dimensional, in contrast to those in California and Japan.

**Acknowledgements**

The work of SA was supported in part by a Grant-in-Aid for Scientific Research from the Japan Society for the Promotion of Science (grant number 26400391). NS acknowledges the support by a Grant-in-Aid for Fundamental Scientific Research from College of Science and Technology, Nihon University.

$(L_{LAT}L_{LON})^{1/2} = 2348.86$ km and $(L_{LAT}L_{LON}L_{DEP})^{1/3} = 583.45$ km. Also, the numerator on the right-hand side of Eq. (5) should be divided by 2.

# Figure Caption

**Fig. 1.** Dependence of the clustering coefficient, $C$, on the dimensionless cell size, $l_3$, (●: California, : Japan, ×: Iran). All quantities are dimensionless.

**Fig. 2.** Dependence of the clustering coefficient, $C$, on the dimensionless cell size, $l = l_3$ for California and Japan, but $l = l_2$ for Iran (●: California, : Japan, ×: Iran). All quantities are dimensionless.

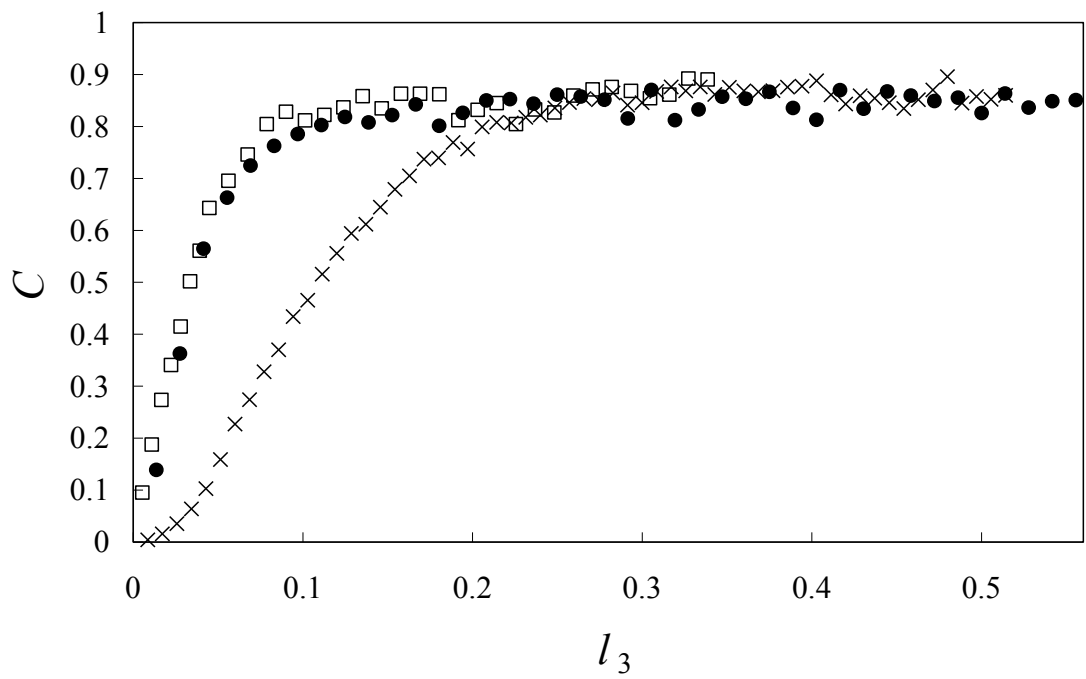

**Fig. 1**

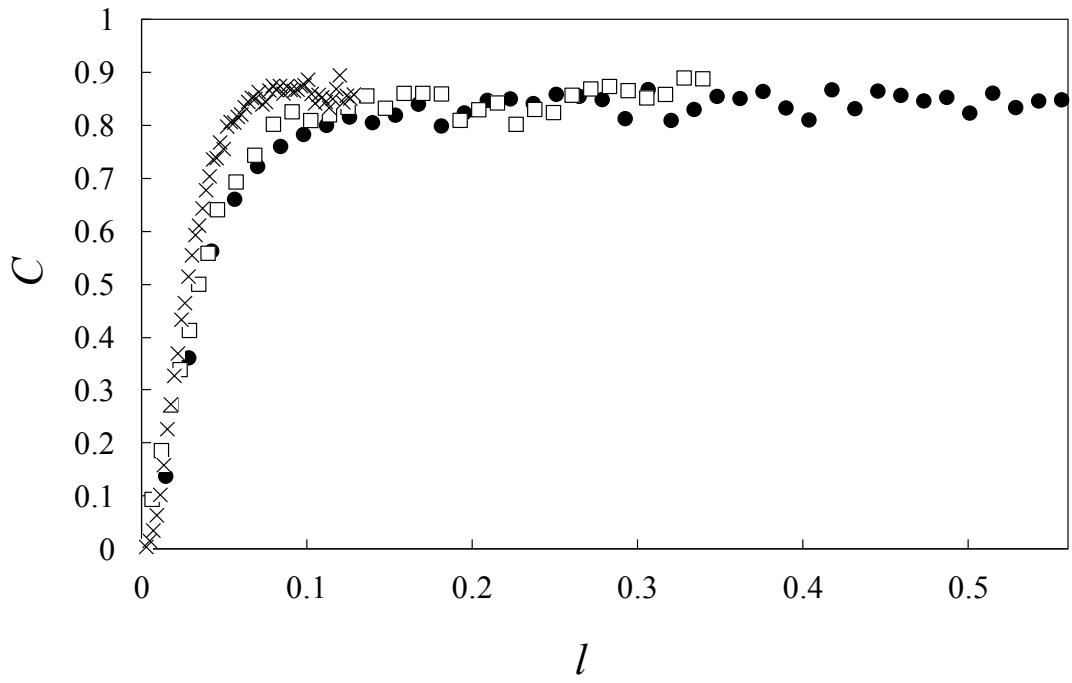

**Fig. 2**